\pdfoutput=1
\documentclass[11pt,twoside,a4paper,cmspaper,final,collab]{cms-tdr}
\def\svnVersion{3aa0eed}\def\svnDate{2024/06/04}\def\cmsCernNoTag{CERN-EP-2023-301}\def\cmsCernDate{\today}\def\cmsMessage{Published in Physical Review D as \href{http://dx.doi.org/10.1103/PhysRevD.109.112003}{\doi{10.1103/PhysRevD.109.112003}.}}
\begin{document}\cmsNoteHeader{EXO-21-019}

\ifthenelse{\boolean{cms@external}}{\providecommand{\cmsLeft}{upper\xspace}}{\providecommand{\cmsLeft}{left\xspace}}
\ifthenelse{\boolean{cms@external}}{\providecommand{\cmsRight}{lower\xspace}}{\providecommand{\cmsRight}{right\xspace}}
\providecommand{\cmsTable}[1]{\resizebox{\textwidth}{!}{#1}}
\newcommand{\mumujj}{\ensuremath{\PGm\PGm\mathrm{jj}}}
\newcommand{\mumubb}{\ensuremath{\PGm\PGm\PQb\PQb}}
\newcommand{\st}{\ensuremath{S_{\mathrm{T}}^{\PGm\PGm\mathrm{jj}}}}
\newcommand{\Muu}{\ensuremath{m_{\PGm\PGm}}}
\newcommand{\Mujone}{\ensuremath{m_{\PGm\mathrm{j1}}}}
\newcommand{\Mujtwo}{\ensuremath{m_{\PGm\mathrm{j2}}}}
\newcommand{\mlq}{\ensuremath{m_{\mathrm{LQ}}\xspace}}
\newcommand{\ttV}{\ensuremath{\ttbar\PV}}
\newcommand{\zjets}{\ensuremath{\PZ/\gamma^*\text{+jets}}}
\newlength\cmsFigWidth
\ifthenelse{\boolean{cms@external}}{\setlength{\cmsFigWidth}{\columnwidth}}{\setlength{\cmsFigWidth}{0.55\textwidth}}

\cmsNoteHeader{EXO-21-019}

\title{Search for pair production of scalar and vector leptoquarks decaying to muons and bottom quarks in proton-proton collisions at \texorpdfstring{$\sqrt{s}=13\TeV$}{sqrt(s) = 13 TeV}}

\author*[cern]{D. M. Morse}

\date{\today}

\abstract{A search for pair production of scalar and vector leptoquarks (LQs) each decaying to a muon and a bottom quark is performed using proton-proton collision data collected at $\sqrt{s}=13\TeV$ with the CMS detector at the CERN LHC, corresponding to an integrated luminosity of 138\fbinv. No excess above standard model expectation is observed. Scalar (vector) LQs with masses less than 1810 (2120)\GeV are excluded at 95\% confidence level, assuming a 100\% branching fraction of the LQ decaying to a muon and a bottom quark. These limits represent the most stringent to date.}

\hypersetup{
pdfauthor={CMS Collaboration},
pdftitle={Search for pair production of scalar and vector leptoquarks decaying to muons and bottom quarks in proton-proton collisions at sqrt(s) = 13 TeV)},
pdfsubject={CMS},
pdfkeywords={CMS, muon, leptoquark, bottom quark}}

\maketitle

\section{Introduction}
\label{Introduction}
\begin{figure*}[ht]
       \centering
       {\includegraphics[width=.25\textwidth]{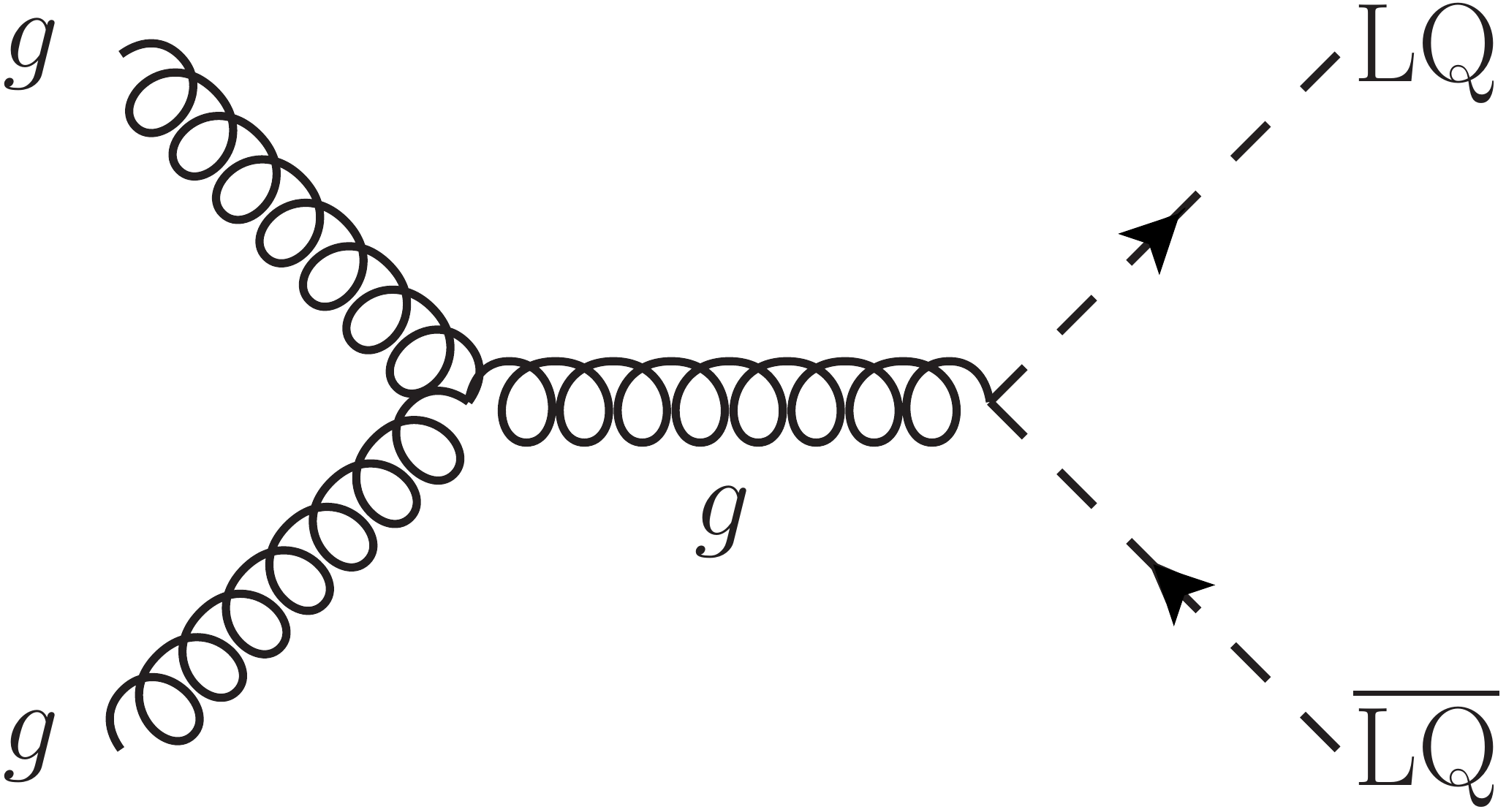}}\hspace*{0.4cm}
       {\includegraphics[width=.25\textwidth]{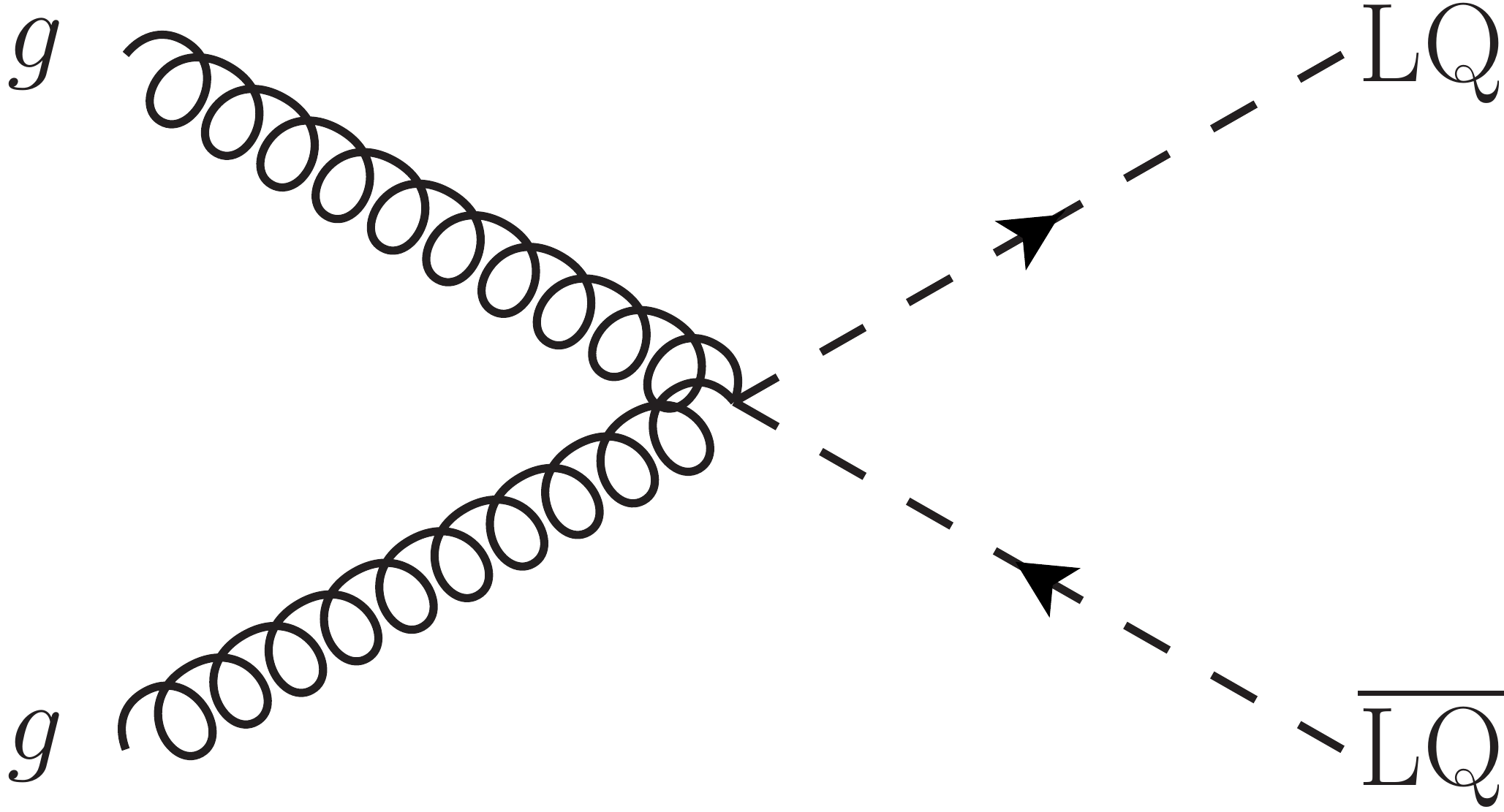}}\hspace*{0.4cm}
       {\includegraphics[width=.25\textwidth]{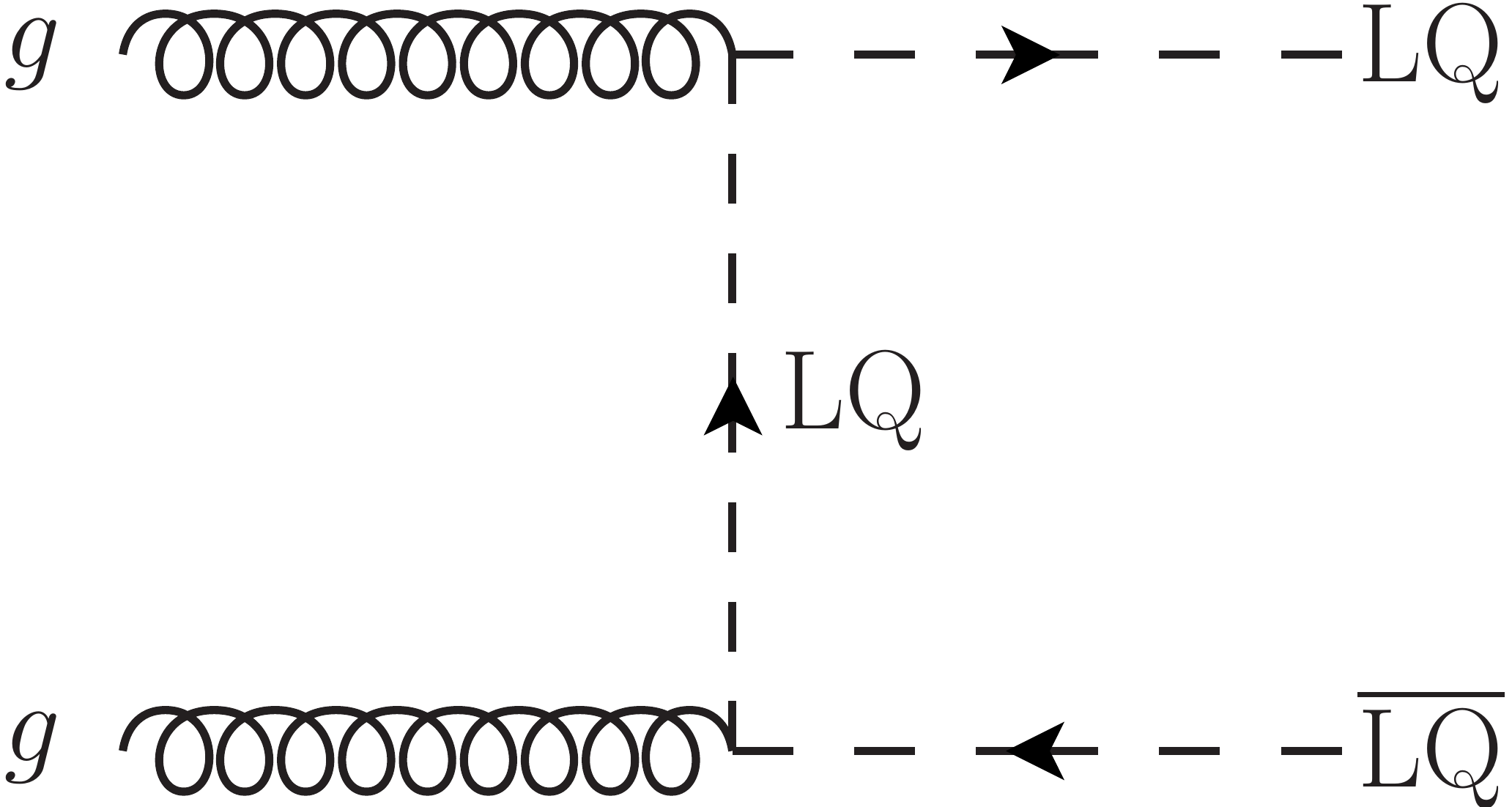}}\\
       \vspace{0.5cm}
       {\includegraphics[width=.25\textwidth]{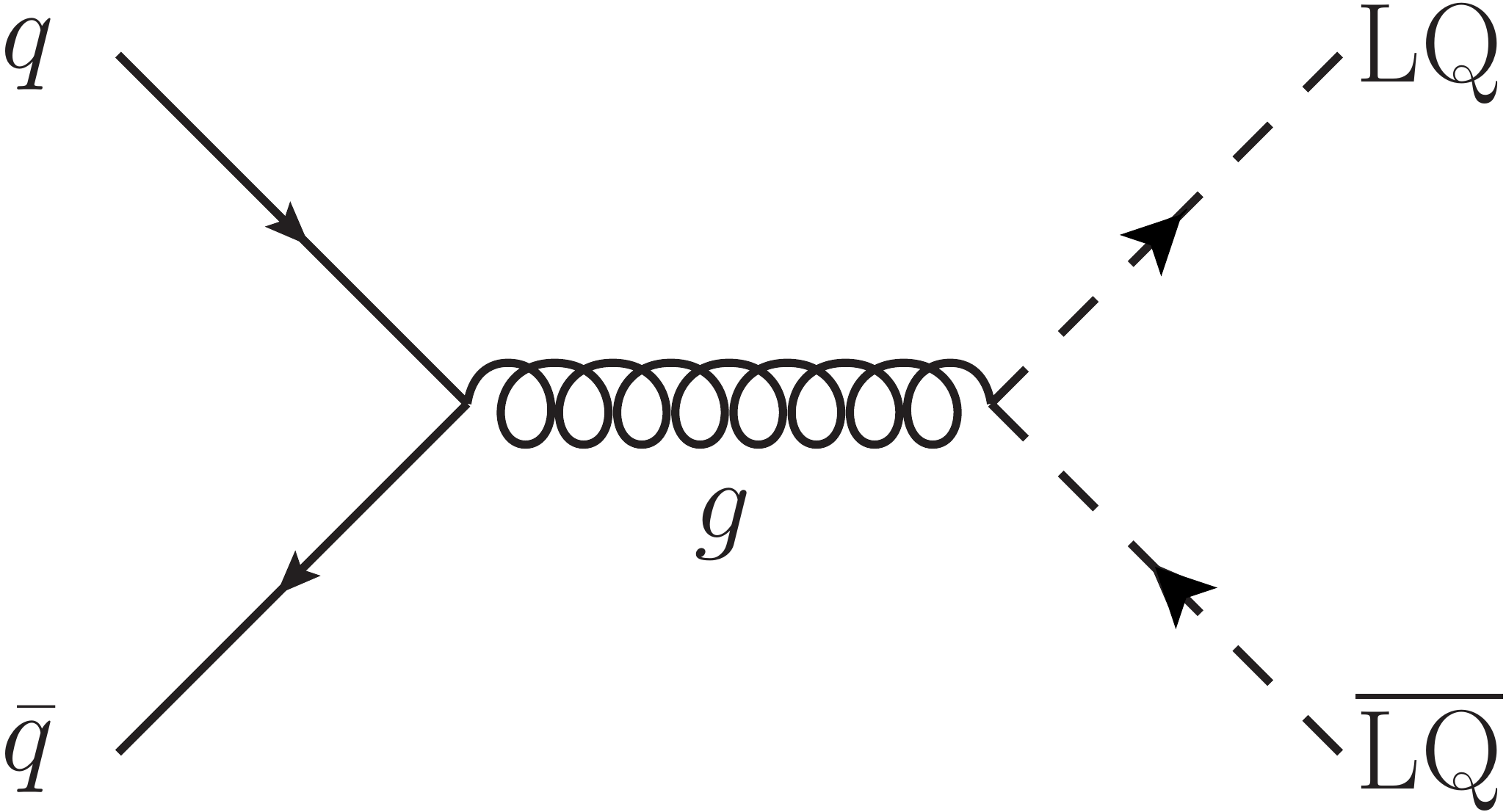}}
       \caption{Dominant leading order Feynman diagrams for pair production of LQs at the LHC.}
          \label{fig:LQFeynman}
\end{figure*}

The standard model (SM) of particle physics describes the interactions of elementary particles to high accuracy. There remain, however, observations of particle physics that are not fully explained, such as the similarities between the quark and lepton families. In some theoretical models of physics with extensions beyond the SM, these similarities are a manifestation of a deeper symmetry between the two types of particles, and there naturally arise particles that bridge the two families. Leptoquarks (LQs) are new bosons that would manifest this fundamental connection between quarks and leptons and are predicted by numerous extensions of the SM, such as grand unified theories~\cite{gut0,gut1,gut2,gut4,gut5,gut6,gut7,gut3}, composite models with lepton and quark substructure~\cite{composite}, technicolor models~\cite{technicolor1,technicolor2,technicolor3}, and superstring-inspired models~\cite{superstring_e6}. Particles with similar decay modes are also found in $R$-parity violating supersymmetry~\cite{Ramond:1971gb,Golfand:1971iw,Neveu:1971rx,Volkov:1972jx,Wess:1973kz,Wess:1974tw,Fayet:1974pd,Farrar:1978xj,Nilles:1983ge,Barbier:2004ez}. Leptoquarks are color-triplet scalar or vector bosons carrying both lepton and baryon numbers. They generally decay either to a charged lepton and a quark, or to a neutrino and a quark. In recent years, there have been a number of observed tensions with the SM in measurements of B meson decays and tests of lepton universality~\cite{BaBar:2012obs,BaBar:2013mob,Belle:2015qfa,Belle:2016dyj,Belle:2017ilt,Belle:2019rba,LHCb:2015gmp,LHCb:2017smo,LHCb:2017rln,LHCb:2017vlu,LHCb:2023cjr,LHCb:2023zxo}, often present in measurements of processes involving muons and third-generation quarks, as well as in precision measurements of the anomalous magnetic moment of the muon~\cite{g-2bnl,g-2fnal,Muong-2:2023cdq}. These tensions have increased interest in LQs and significant theoretical literature~\cite{Dorsner:2016wpm,Diaz:2017lit,Buttazzo:2017ixm,Calibbi:2017qbu,Hiller:2018wbv} has re-examined the phenomenological and historical constraints of LQ models and searches, in particular removing the traditional constraint that LQs should decay within the same generation of quarks and leptons. Leptoquarks may provide a tree-level explanation of some anomalies in scenarios where the LQ-lepton-quark coupling is not restricted to remain in one generation~\cite{Diaz:2017lit}. At hadron colliders, LQs can be produced singly or in pairs. The dominant leading-order (LO) processes for pair production of LQs at the LHC involve gluon-gluon fusion and quark-antiquark annihilation, shown in Fig.~\ref{fig:LQFeynman}. Pair production can also arise from so-called t-channel production, which does not provide a significant contribution for the model and parameters chosen in this analysis and is not considered~\cite{Dorsner:2014axa}.

Interpretations of direct searches for LQs are typically based on a general model where LQ-lepton-quark interactions are added to the SM Lagrangian~\cite{mBRW}. The interactions of scalar LQs with SM particles are completely determined by three parameters~\cite{mBRW}; the LQ mass {\mlq}, the Yukawa coupling at the LQ-lepton-quark vertex $\lambda_{\mathrm{LQ}}$, and the branching fraction $\beta$ of the LQ decay to a charged lepton and a quark. Vector LQs are further dependent on an additional parameter $\kappa$ that relates to their anomalous magnetic and electric quadrupole moments~\cite{Blumlein:1996qp}. Values of $\kappa = 0$ and 1 are considered in this analysis, corresponding to the minimal coupling and Yang--Mills scenarios, respectively. Leptoquark pair production cross sections ($\sigma$) are independent of $\lambda_{\mathrm{LQ}}$ over many orders of magnitude~\cite{mBRW}. In this analysis, $\lambda_{\mathrm{LQ}}$ has been set to 1, which will produce LQs that decay very close to the point of production~\cite{Diaz:2017lit} and ensures the pair production cross sections and results are independent of $\lambda_{\mathrm{LQ}}$.

Pair production of LQs is characterized by final states with two leptons and two jets with large transverse momentum \pt. Previous limits on scalar LQ pair production decaying to muons and jets in proton-proton ($\Pp\Pp$) collisions have been published by the CMS and ATLAS Collaborations~\cite{Khachatryan:2015vaa,PhysRevD.99.032014,ATLAS:2020dsk,ATLAS:2022wcu,ATLAScomb}. The CMS result excludes, at 95\% confidence level (\CL), LQs decaying to muons and light (u, d, s) quarks with $\mlq < 1.5$\,(1.3)\TeV for $\beta = 1$\,$(0.5)$~\cite{PhysRevD.99.032014}, and ATLAS excludes LQs decaying to muons and {\PQb} quarks with $\mlq < 1.7\TeV$ for $\beta=1$~\cite{ATLAS:2022wcu,ATLAScomb}. Previous limits on vector LQs have been reported by CMS~\cite{Khachatryan:2015vaa} and ATLAS~\cite{ATLAScomb}, where CMS excludes vector LQs decaying to muons and light quarks with $\mlq < 1.3$ (1.5)\TeV for $\beta=1$ in the minimal coupling (Yang--Mills) scenario, and ATLAS excludes vector LQs decaying to muons and {\PQb} quarks with $\mlq < 1.7$ (2.0)\TeV for $\beta=0.5$ in the minimal coupling (Yang--Mills) scenario. The present analysis searches for scalar and vector LQs that decay to a muon and a bottom quark, using data collected by the CMS detector during the 2016--2018 {$\Pp\Pp$} LHC runs at $\sqrt{s} = 13\TeV$. Results are presented for a value of $\beta=1$, corresponding to maximal production of the {\mumubb} final state, as well as in the range $0<\beta<1$.

\section{The CMS detector}
\label{CMS}
The central feature of the CMS apparatus is a superconducting solenoid of 6\unit{m} internal diameter, providing a magnetic field of 3.8\unit{T}. Within the solenoid volume are a silicon pixel and strip tracker, a lead tungstate crystal electromagnetic calorimeter (ECAL), and a brass and scintillator hadron calorimeter (HCAL), each composed of a barrel and two endcap sections. Forward calorimeters extend the pseudorapidity ($\eta$) coverage provided by the barrel and endcap detectors. Muons are measured in gas-ionization detectors embedded in the steel flux-return yoke outside the solenoid. A more detailed description of the CMS detector, together with a definition of the coordinate system used and the relevant kinematic variables, can be found in Ref.~\cite{CMS:2008xjf}.

Events of interest are selected using a two-tiered trigger system. The first level (L1), composed of custom hardware processors, uses information from the calorimeters and muon detectors to select events at a rate of around 100\unit{kHz} within a fixed latency of about 4\mus~\cite{CMS:2020cmk}. The second level, known as the high-level trigger, consists of a farm of processors running a version of the full event reconstruction software optimized for fast processing, and reduces the event rate to around 1\unit{kHz} before data storage~\cite{CMS:2016ngn}.

\section{Event reconstruction}
\label{selection}
 The primary vertex (PV) is taken to be the vertex corresponding to the hardest scattering in the event, evaluated using tracking information alone, as described in Section 9.4.1 of Ref.~\cite{CMS-TDR-15-02}. The global event reconstruction (also called particle-flow event reconstruction~\cite{CMS:2017yfk}) aims to reconstruct and identify each individual particle in an event, with an optimized combination of all subdetector information. In this process, the identification of the particle type (photon, electron, muon, charged or neutral hadron) plays an important role in the determination of the particle direction and energy. Photons (\eg, coming from \PGpz decays or from electron bremsstrahlung) are identified as ECAL energy clusters not linked to the extrapolation of any charged particle trajectory to the ECAL. Electrons (\eg, coming from photon conversions in the tracker material or from \PB hadron semileptonic decays) are identified as primary charged particle tracks and potentially many ECAL energy clusters corresponding to these track extrapolations to the ECAL and to possible bremsstrahlung photons emitted along the way through the tracker material. Muons (\eg, from \PB hadron semileptonic decays) are identified as tracks in the central tracker consistent with either a track or several hits in the muon system, and associated with calorimeter deposits compatible with the muon hypothesis. Charged hadrons are identified as charged particle tracks neither identified as electrons, nor as muons. Finally, neutral hadrons are identified as HCAL energy clusters not linked to any charged hadron trajectory, or as a combined ECAL and HCAL energy excess with respect to the expected charged hadron energy deposit.

Jets originating from bottom quarks are identified using the \textsc{DeepJet} discriminator~\cite{Bols:2020bkb}, a multiclass jet flavor tagging algorithm that uses deep learning and exploits low-level features from a high number of jet constituents. A working point with a 75\% efficiency to identify {\PQb} jets and a 1\% misidentification probability for light jets is used in this analysis~\cite{BTV-16-002,CMS-DP-2023-005}.

The energy of photons is obtained from the ECAL measurement. The energy of electrons is determined from a combination of the track momentum at the main interaction vertex, the corresponding ECAL cluster energy, and the energy sum of all bremsstrahlung photons attached to the track. The energy of muons is obtained from the corresponding track momentum. The energy of charged hadrons is determined from a combination of the track momentum and the corresponding ECAL and HCAL energy deposits, corrected for the response function of the calorimeters to hadronic showers. Finally, the energy of neutral hadrons is obtained from the corresponding corrected ECAL and HCAL energy deposits.

For each event, hadronic jets are clustered from these reconstructed particles using the infrared and collinear safe anti-\kt algorithm~\cite{Cacciari:2008gp, Cacciari:2011ma} with a distance parameter of 0.4. Jet momentum is determined as the vectorial sum of all particle momenta in the jet, and is found from Monte Carlo (MC) simulation to be, on average, within 5--10\% of the true momentum over the whole \pt spectrum and detector acceptance. Additional {$\Pp\Pp$} interactions within the same or nearby bunch crossings (pileup) can contribute additional tracks and calorimetric energy depositions to the jet momentum. To mitigate this effect, charged particles identified to be originating from pileup vertices are discarded and an offset correction is applied to correct for remaining contributions~\cite{CMS:2020ebo}. Jet energy corrections are derived from simulation to bring the measured response of jets to that of particle level jets on average. In situ measurements of the momentum balance in dijet, $\text{photon+jet}$, $\PZ\text{+jet}$, and multijet events are used to account for any residual differences in the jet energy scale between data and simulation~\cite{CMS:2016lmd}. The jet energy resolution amounts typically to 15--20\% at 30\GeV, 10\% at 100\GeV, and 5\% at 1\TeV~\cite{CMS:2016lmd}. Additional selection criteria are applied to each jet to remove jets potentially dominated by anomalous contributions from various subdetector components or reconstruction failures~\cite{CMS:2020ebo}.

Muons are measured in the range $\abs{\eta} < 2.4$, with detection planes made using three technologies: drift tubes, cathode strip chambers, and resistive plate chambers. The single muon trigger efficiency exceeds 90\% over the full $\eta$ range, and the efficiency to reconstruct and identify muons is greater than 96\%. Matching muons to tracks measured in the silicon tracker results in a relative \pt resolution of 1\% in the barrel and 3\% in the endcaps for muons with \pt up to 100\GeV, and of better than 7\% in the barrel for muons with \pt up to 1\TeV~\cite{CMS:2018rym}. Muons are required to be isolated from tracks arising from electromagnetic and hadronic activity in the event that may otherwise mimic muon signatures in the detector.

The missing transverse momentum vector \ptvecmiss is computed as the negative vector \pt sum of all the particle-flow candidates in an event, and its magnitude is denoted as \ptmiss~\cite{CMS:2019ctu}. The \ptvecmiss is modified to account for corrections to the energy scale of the reconstructed jets in the event.

\section{Data and simulated samples}
\label{samples}

The data set used in this paper was collected by the CMS detector during the 2016--2018 {$\Pp\Pp$} LHC runs at $\sqrt{s} = 13\TeV$ and corresponds to an integrated luminosity of $138\fbinv$~\cite{LUM-17-003,LUM-17-004,LUM-18-002}. The integrated luminosities for the 2016, 2017, and 2018 data-taking years have 1.2--2.5\% individual uncertainties~\cite{LUM-17-003,LUM-17-004,LUM-18-002}, while the overall uncertainty for the 2016--2018 period is 1.6\%. Events are selected using triggers that require at least one muon with $\pt>50\GeV$, with no isolation requirements. 

Signal samples are produced in 100\GeV steps for values of {\mlq} between 300--3000\GeV at LO in quantum chromodynamics (QCD) with {\MGvATNLO}~\cite{amcnlo,MadGraph} versions 2.2.2 and 2.3.3 using the triplet scalar S3 model of Ref.~\cite{Dorsner:2016wpm}. For this analysis, the LQ ($\overline{\text{LQ}}$) is forced to decay 100\% to $\text{LQ}\to\PGm^-\PQb\ $ ($\overline{\text{LQ}}\to\PGm^+\overline{\PQb}$), giving rise to a $\PGm^-\PGm^+\PQb\overline{\PQb}$ final state. These samples are used to study the signal efficiency.

\sloppy The main SM background processes in this final state are {\ttbar}+jets and {\zjets} production. Other minor contributions from SM processes include diboson ($\PW\PW$/$\PW\PZ$/$\PZ\PZ$)+jets, $\ttbar+\PZ$/$\PW$/$\PH$ (\ttV), single top quark production, and $\PW$+jets. Background from QCD multijets has been shown to be negligible~\cite{PhysRevD.99.032014} and is not considered in this analysis.

The {\zjets} and $\PW$+jets samples are generated at next-to-LO (NLO) using {\MGvATNLO} with \textsc{FxFx} merging~\cite{Frederix:2012ps}. Diboson samples are generated at NLO using {\MGvATNLO}. Single top quark, {\ttbar}+jets, and {\ttV} samples are generated at NLO using \textsc{powheg} v2~\cite{Frixione:2007vw,powheg,Alioli:2009je,Alioli:2010xd} and {\MGvATNLO} \cite{Frixione:2005vw}. All of these samples are showered using \textsc{pythia} 8.212~\cite{pythia8p2}, to take into account the fragmentation and hadronization processes.

The parton distribution function (PDF) sets used for generating the 2016 (2017--2018) signal and background samples are \textsc{nnpdf3.0} NLO (3.1 LO for signal, 3.1 NNLO for background)~\cite{nnpdf}, with the full CMS detector geometry and response simulated using {\GEANTfour}~\cite{Geant1,Geant2}; all samples use the \textsc{cuetp8m1}~\cite{Khachatryan:2015pea} (\textsc{cp2} for signal, \textsc{cp5} for SM backgrounds)~\cite{CMS:2019csb}) underlying event tune, with additional {$\Pp\Pp$} interactions overlaid and corrected to match the distribution measured in data.

The scalar LQ pair production cross sections are calculated with {\MGvATNLO} using the tool from Ref.~\cite{lqToolbox} with NLO QCD corrections~\cite{kramer,PhysRevD.93.035018,lqToolbox} and the \textsc{PDF4LHC15}~\cite{pdf-syst-1} NLO PDF set, and are used for comparison with data and background estimations in the statistical analysis. For the vector LQ interpretation, LQ production cross sections are calculated at LO in QCD using a singlet U1 model in which the LQ couples only to left-handed fermions~\cite{lqToolbox}. The scalar LQ simulated samples are again used for the signal efficiency. It has been shown that relevant vector LQ pair production kinematic distributions agree within 10\% with equivalent scalar pair production, for $\mlq>500\GeV$~\cite{Diaz:2017lit}.

The simulated samples are corrected so that the detector response and resolution for both leptons and jets (including \PQb tagged jets) and the trigger efficiency match those measured in data~\cite{CMS:2016lmd,CMS:2018rym}.

\section{Background estimation}
\label{bg}
\begin{figure*}[htb!]
       \centering
       {\includegraphics[width=.47\textwidth]{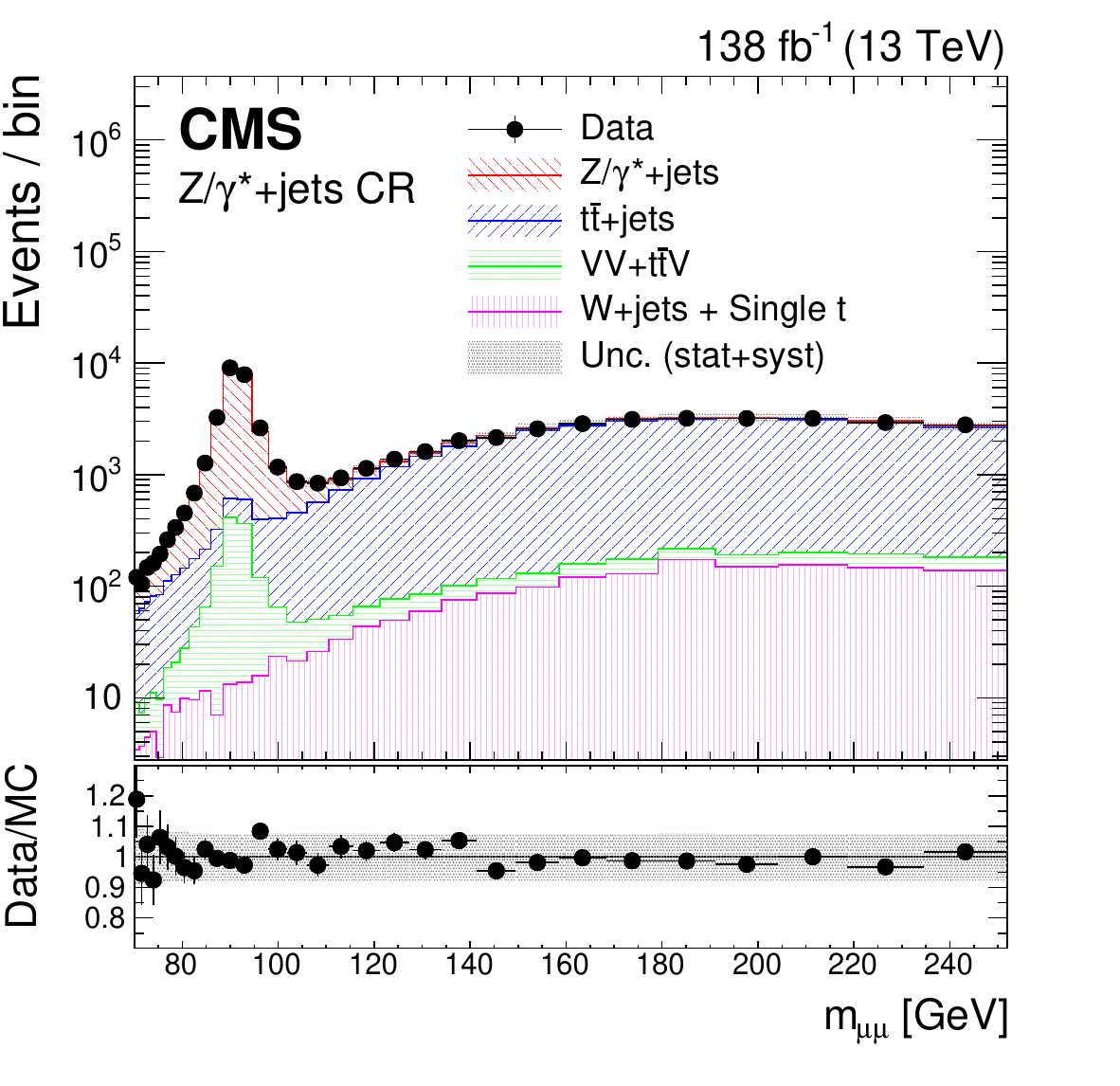}}
       {\includegraphics[width=.47\textwidth]{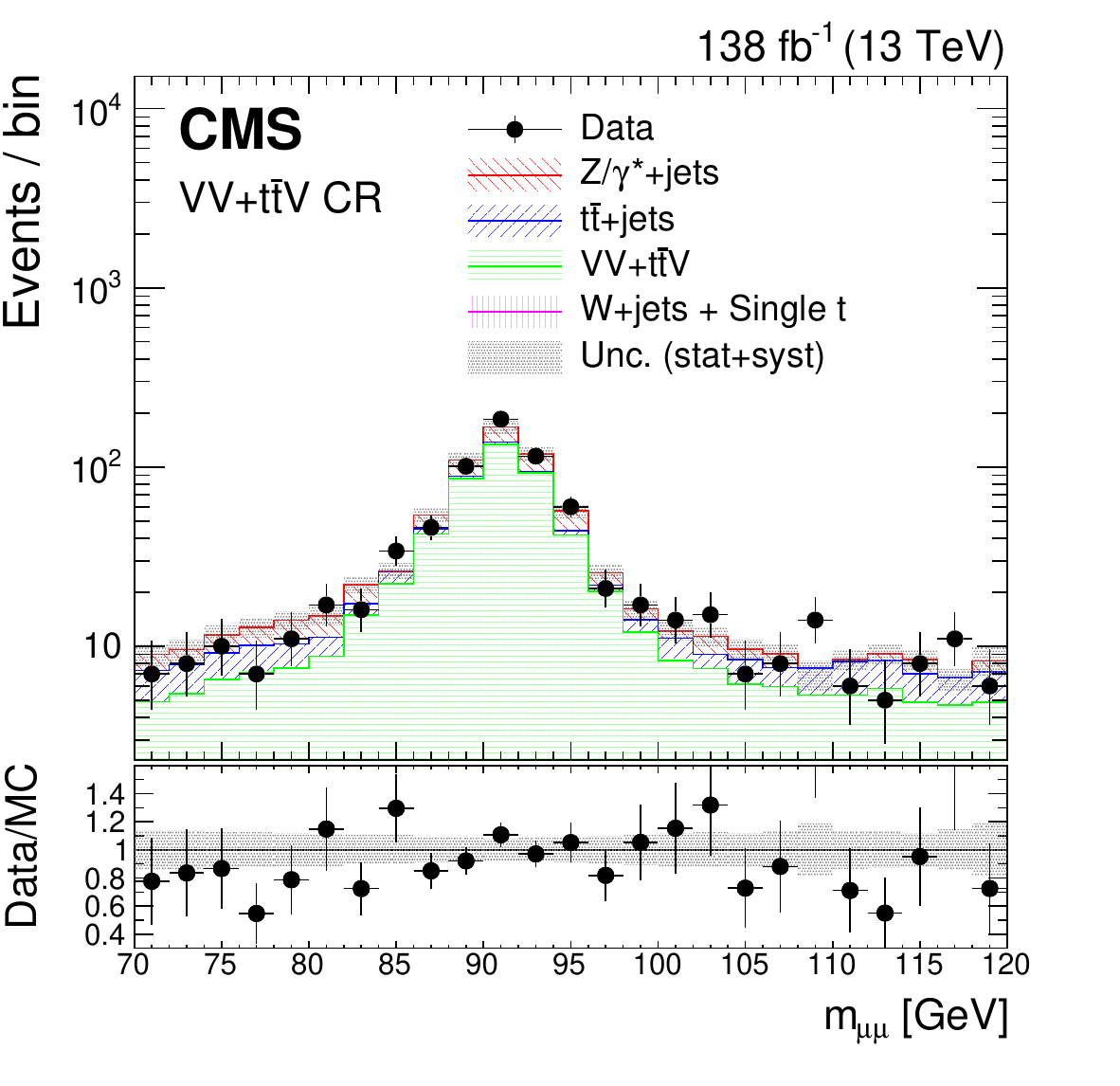}}
       \caption{Comparison of data and background $\Muu$ distribution at the preselection level for the {\zjets} and {\ttbar}+jets (left) and diboson and {\ttV} (right) background control regions, with the corresponding data-to-background ratio shown below. For {\zjets}, the control region is a $\Muu$ window of 80--100\GeV around the {\PZ} peak, and for {\ttbar}+jets is a window of 100--250\GeV.  For diboson and {\ttV} processes, the normalization is performed simultaneously, with a control region again of 80--100\GeV around the {\PZ} peak, but with a third lepton requirement (to remain orthogonal to the {\zjets} control region) and no {\PQb} tag requirement (to diminish the statistical uncertainty). The error bars are the data statistical uncertainties, while the shaded band represents the combined statistical and systematic uncertainty in the full background estimate. The signal contribution in all control regions is negligible.}
          \label{fig:control}
\end{figure*}

The main SM processes that can mimic the LQ signal in this channel are \zjets, {\ttbar}+jets, and to a lesser extent diboson production. Backgrounds are estimated and validated using a selection dominated by background events, referred to as the preselection. This preselection requires two muons with $\pt>53\GeV$, to remain fully efficient with respect to the trigger requirements, and at least two jets with $\pt>50\GeV$, at least one of which is tagged as a bottom jet. Each event is required to have only two muons, and a veto on a third muon or an electron with $\pt>20\GeV$ is imposed. No requirement is placed on the muon charges. The muons are required to be separated from each other by $\Delta R = \sqrt{\smash[b]{(\Delta\eta)^2+(\Delta\phi)^2}}>0.3$, where {$\phi$} is the azimuthal angle (in radians) with respect to the counterclockwise beam axis. There are two requirements of $\Muu>50\GeV$ and $\st>300\GeV$, where $\Muu $is the invariant mass of the dimuon system, and $\st$ is defined as the scalar {\pt} sum of the two leading jets and two muons in the event.

The \zjets, {\ttbar}+jets, diboson, and {\ttV} contribution estimates make use of background-dominated data control regions (CRs) after the preselection. Background shapes are taken from simulation, and show good shape agreement with the data in the CRs. For normalization, the simulation is compared to data in the different CRs, and the measured data normalization scale factors are applied to simulated events in the analysis. The {\zjets} and {\ttbar} normalizations are computed iteratively, then held constant to compute the simultaneous diboson and {\ttV} normalization. For {\zjets}, the CR is a $\Muu$ window of 80--100\GeV around the {\PZ} peak, and for {\ttbar}+jets it is a window of 100--250\GeV. Both are computed at the preselection level. For diboson and {\ttV} processes, the normalization is performed simultaneously, with a CR again $80<\Muu<100\GeV$ around the {\PZ} peak, but with the preselection modified by the additional request of a third lepton (to remain orthogonal to the {\zjets} CR) and no {\PQb} tag requirement (to diminish the statistical uncertainty). Background contributions from single top quark and $\PW$+jets events are estimated from simulation. The CR $\Muu$ distributions, after application of the data normalization scale factors, are shown in Fig.~\ref{fig:control}. The signal contribution in all control regions is negligible.

\begin{figure*}[htb!]
       \centering
       {\includegraphics[width=.45\textwidth]{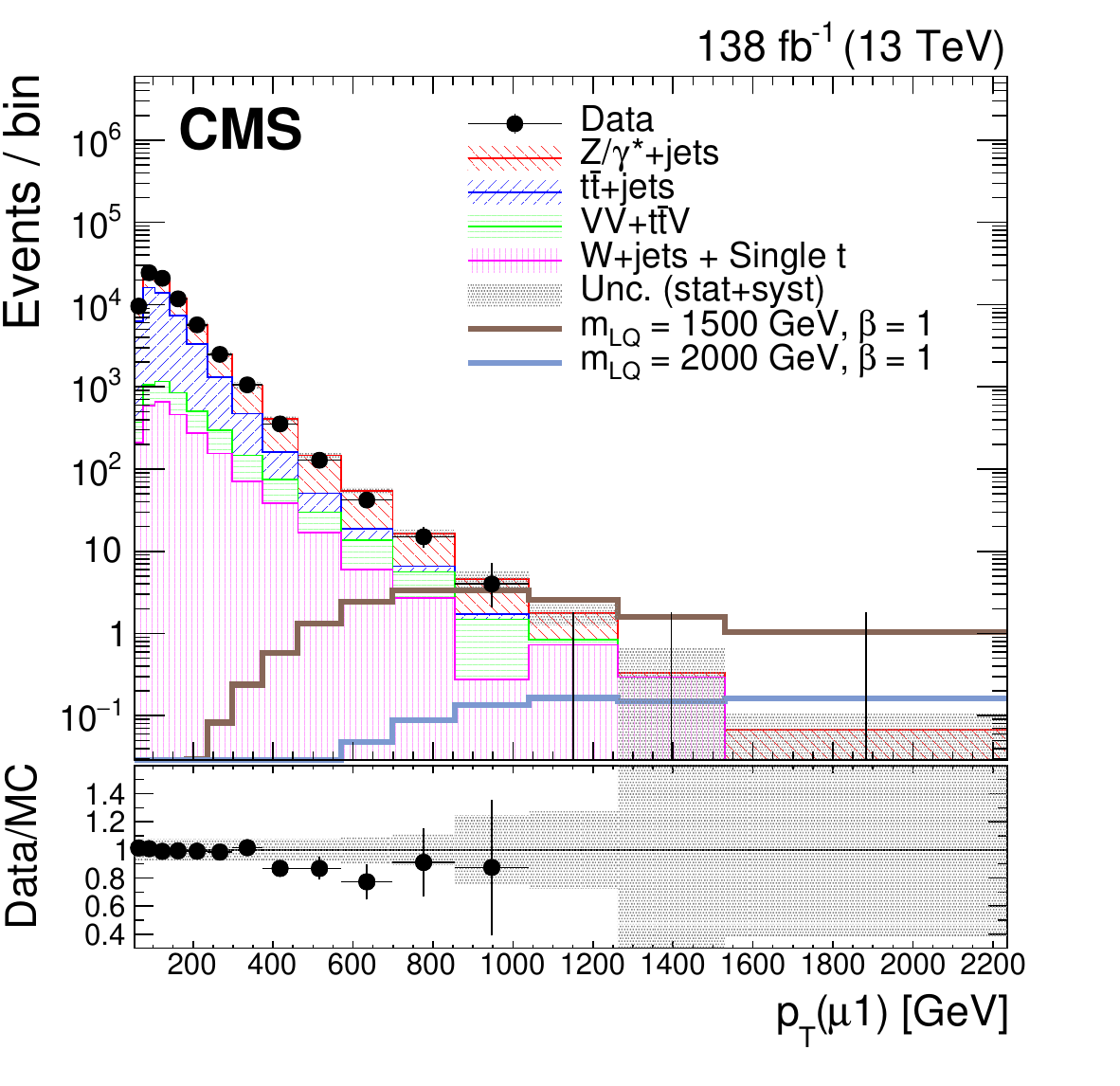}}
       {\includegraphics[width=.45\textwidth]{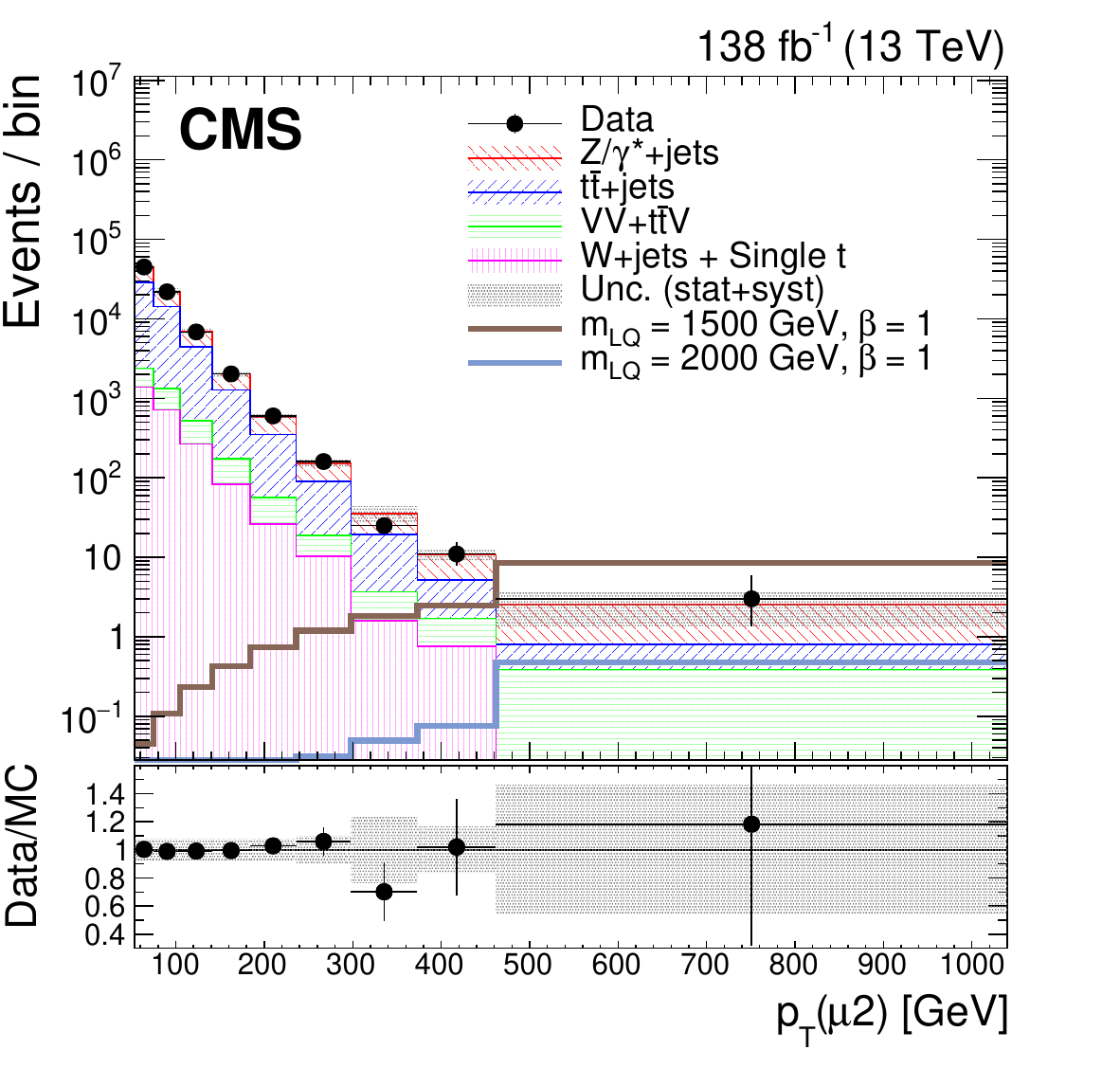}}
       {\includegraphics[width=.45\textwidth]{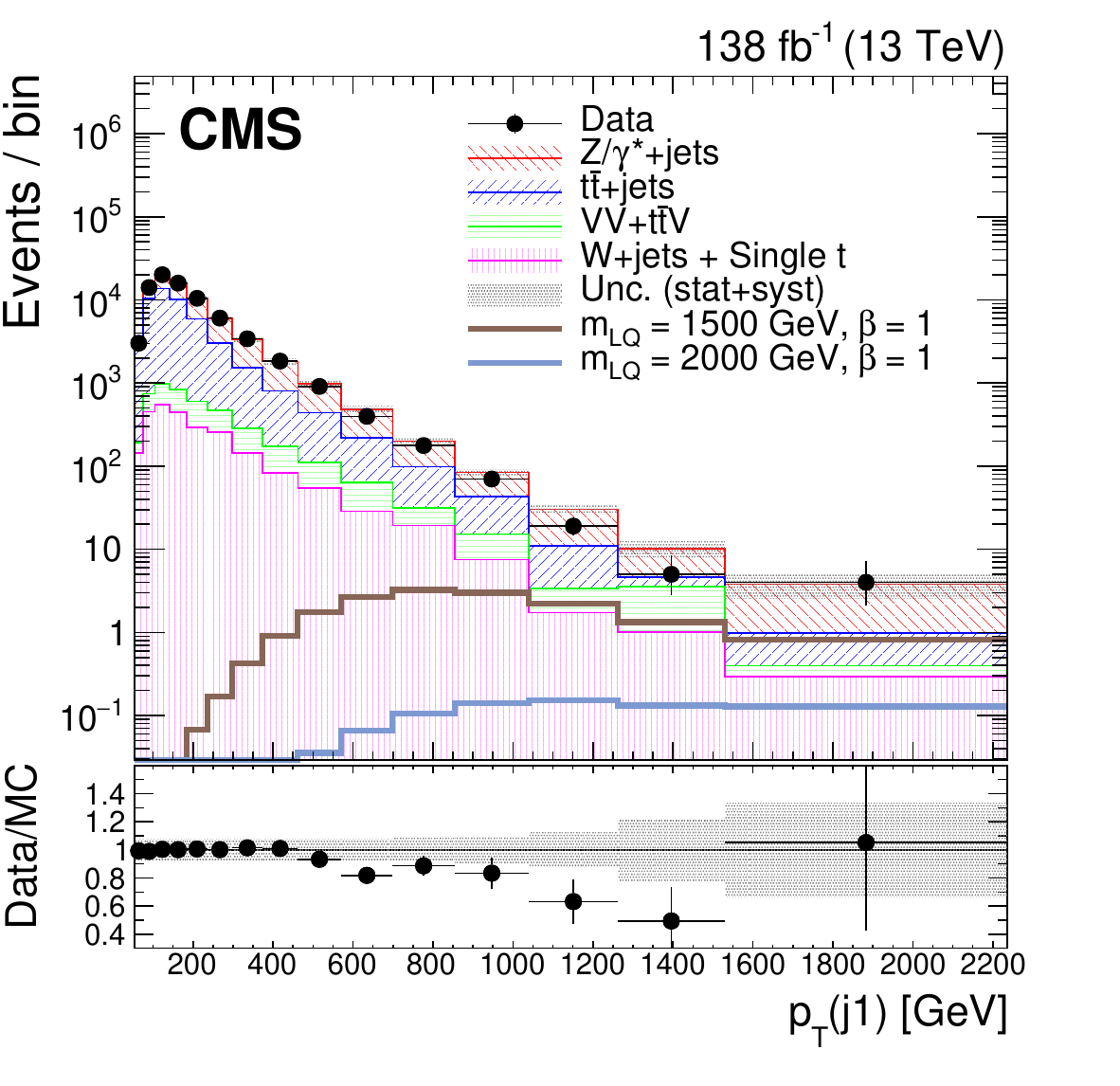}}
       {\includegraphics[width=.45\textwidth]{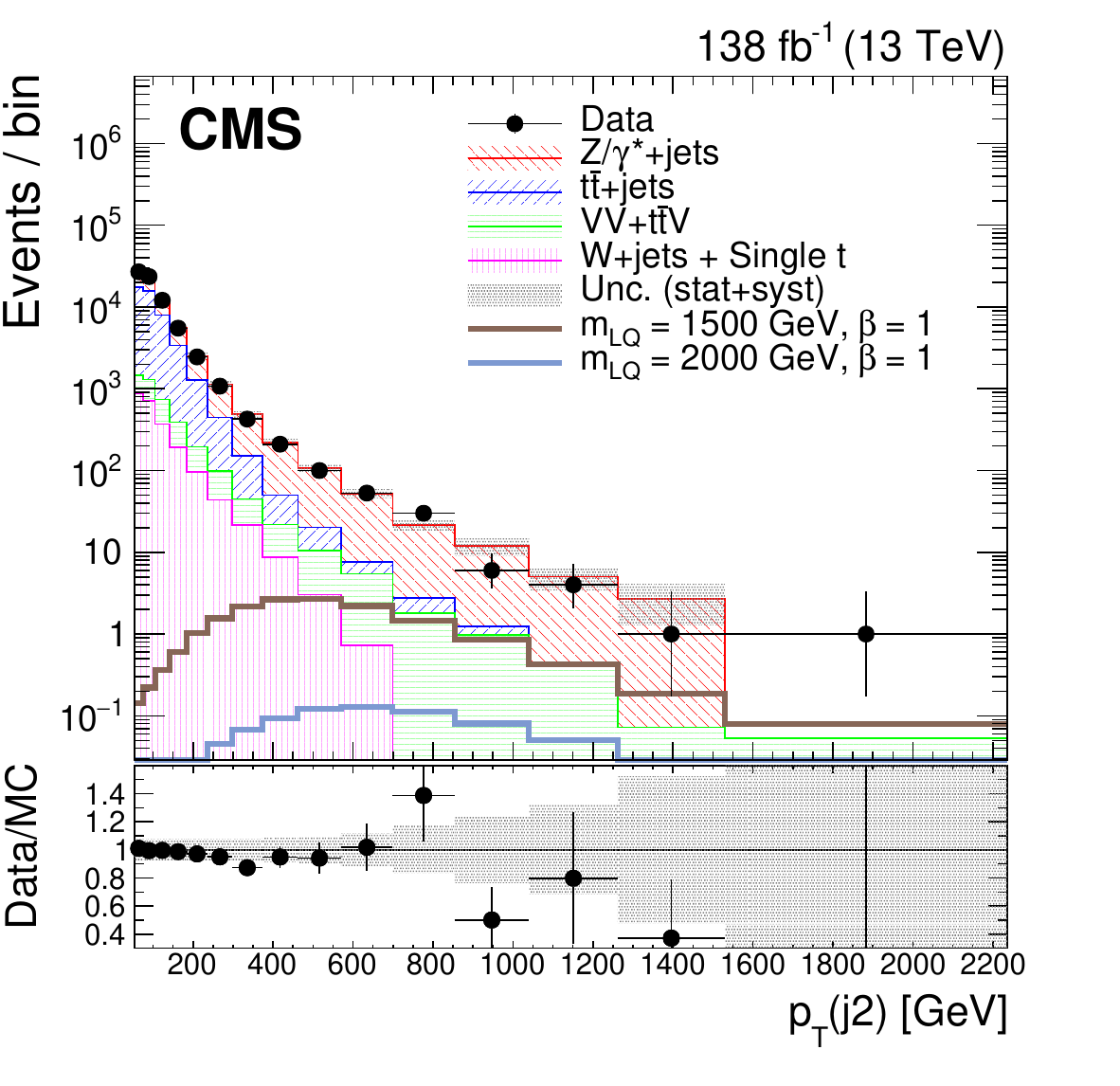}}
       \caption{Comparison of data and background $\pt$ distribution at the preselection level for the leading two muons and jets. The error bars are the data statistical uncertainties, while the shaded band represents the combined statistical and systematic uncertainty in the full background estimate.}
          \label{fig:pts}
\end{figure*}

The background predictions are validated at the preselection level by comparing them with data in all relevant kinematic distributions. Distributions at preselection of the muon and jet {$\pt$} spectra are shown in Fig.~\ref{fig:pts}. There is good agreement apart from in the high {$\pt$} regions of the jet spectra, where there is a tendency for the simulation to overestimate the rate. The small discrepancy is not important since the analysis sensitivity is dependent on the final selections discussed in Section~\ref{finselopt}, and the jet {$\pt$} distributions play a limited role in the final selection.

\section{Results}
\label{results}

\subsection{Final selection optimization}
\label{finselopt}
In order to separate signal from background, a final selection is defined for each {\mlq} hypothesis. Multivariate techniques are used to increase the signal sensitivity compared to an analysis based on a set of independent selection criteria. A boosted decision tree (BDT) is trained on background and signal events at a modified preselection level for each {\mlq} hypothesis.  The modified preselection requires additional selection criteria of $\Muu>250\GeV$ and $m_{\mumujj}>\mlq$, to maintain the separation of the CR from the signal regions and reduce training bias in background-enriched and signal-depleted regions. An iterative procedure identifies eleven minimally-correlated kinematic variables with strong signal-to-background separation power to use as the input variables to the BDTs:

\begin{itemize}
\item invariant masses: $\Muu$ and $m_{\mumujj}$;
\item reconstructed LQ invariant masses $\Mujone$ and $\Mujtwo$, defined by pairing the two muons and jets such as to minimize the mass difference;
\item final-state momenta of muons and jets: $\pt(\PGm1)$, $\pt(\PGm2)$, $\pt(\mathrm{j1})$, and $\pt(\mathrm{j2})$;
\item combined momenta: \st and \ptmiss;
\item $\Delta R$ separation between the dimuon pair and the leading-\pt jet momentum vectors.
\end{itemize}

\begin{figure*}[ht]
       \centering
       {\includegraphics[width=.45\textwidth]{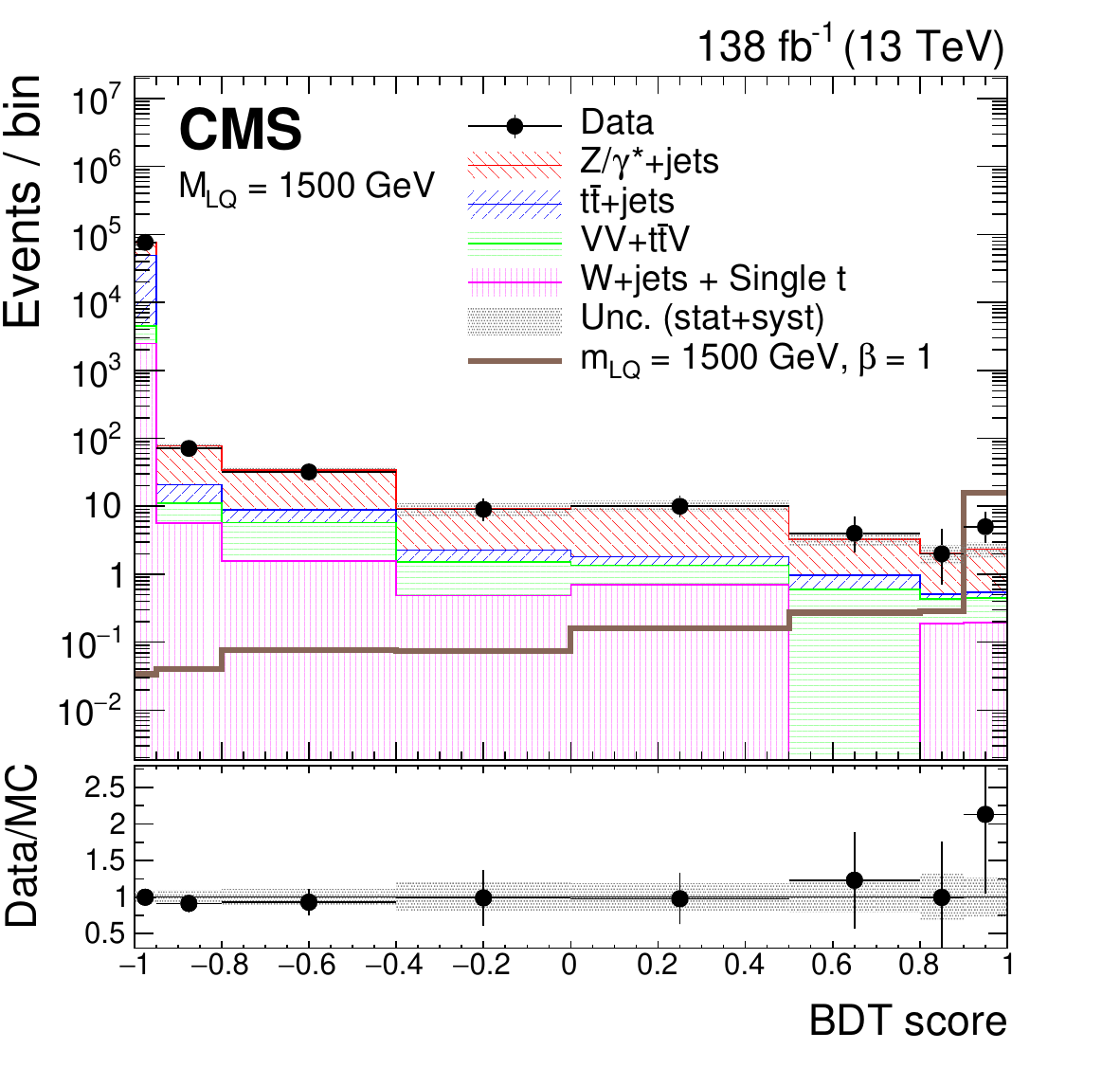}}
       {\includegraphics[width=.45\textwidth]{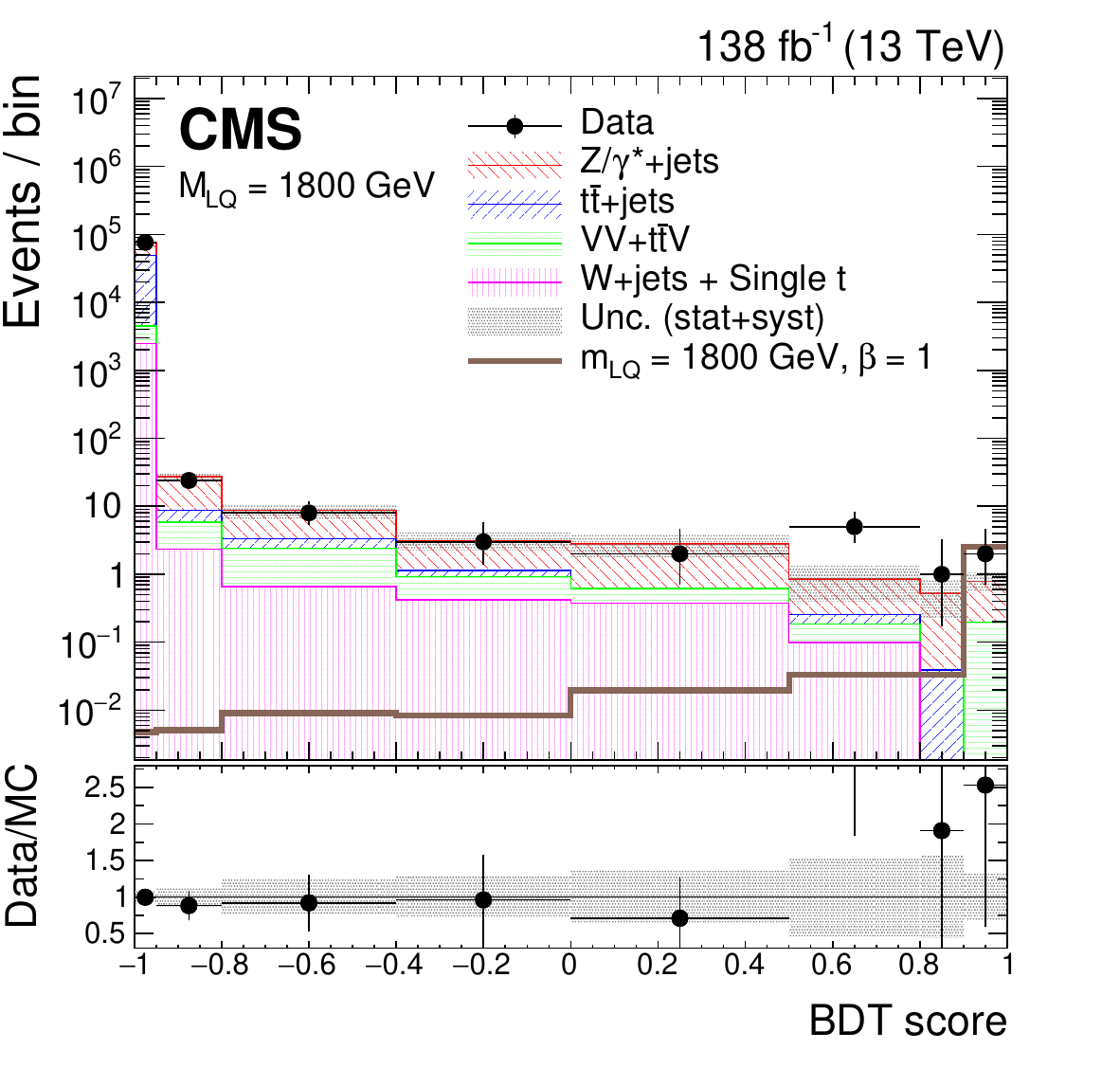}}
       {\includegraphics[width=.45\textwidth]{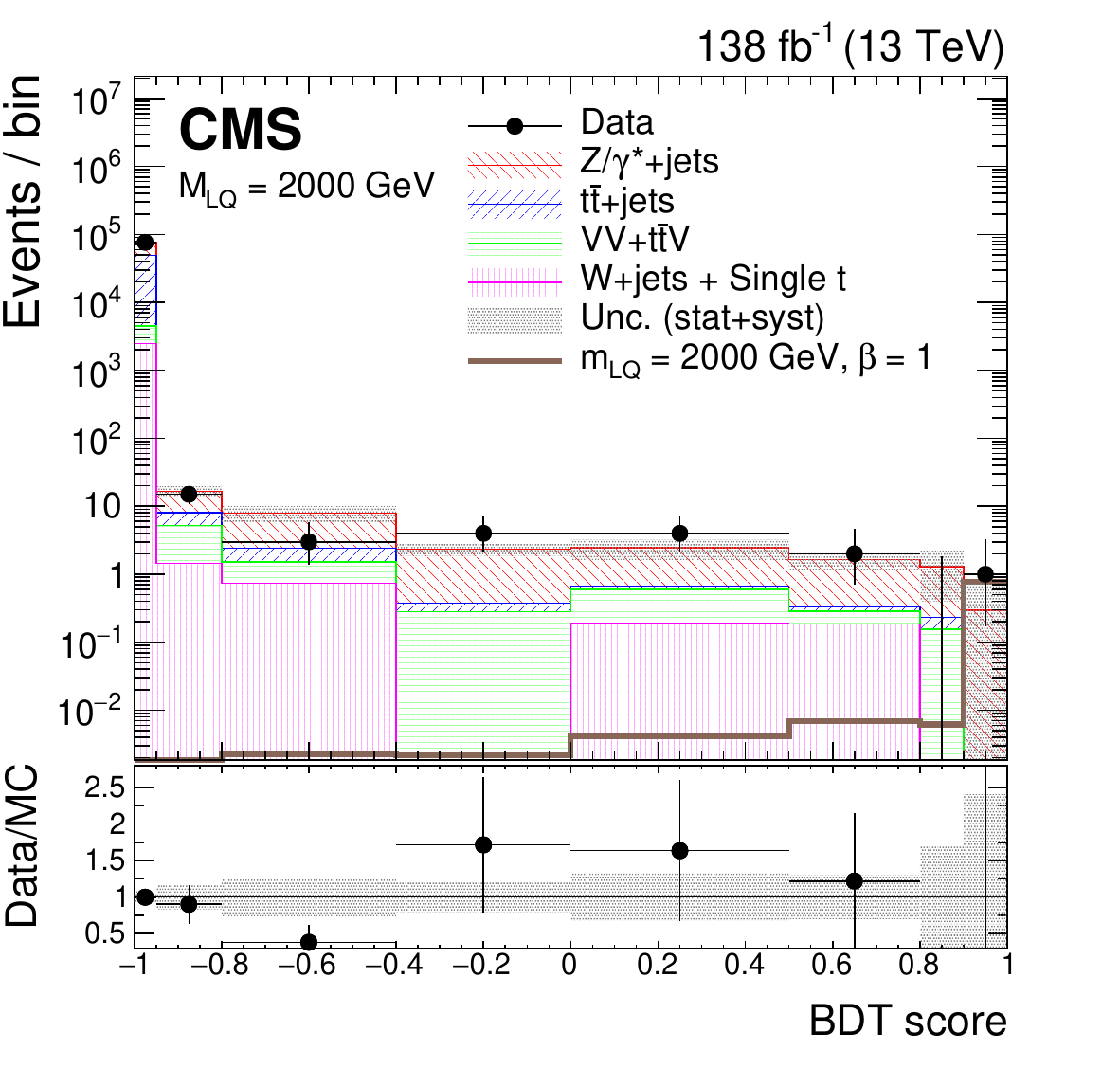}}
       \caption{Comparison of data and background BDT discriminant distributions at the preselection level for LQ mass hypotheses of 1500\GeV (upper left), 1800\GeV (upper right), and 2000\GeV (lower). The error bars are the data statistical uncertainties, while the shaded band represents the combined statistical and systematic uncertainty in the full background estimate.}
          \label{fig:BDT}
\end{figure*}
 
Training samples are constructed using the full MC data set, with random event sampling from the background and signal simulated samples. The gradient boost~\cite{mbbf-bagd-00,friedman2001greedy,grad1} algorithm is used for the BDT training, avoiding input variables that are too highly correlated, and studies of the independent training and testing events show no evidence of overtraining. Examples of the BDT response at the preselection level are shown in Fig.~\ref{fig:BDT}.

\begin{figure*}[ht]
       \centering
       \includegraphics[width=0.75\textwidth]{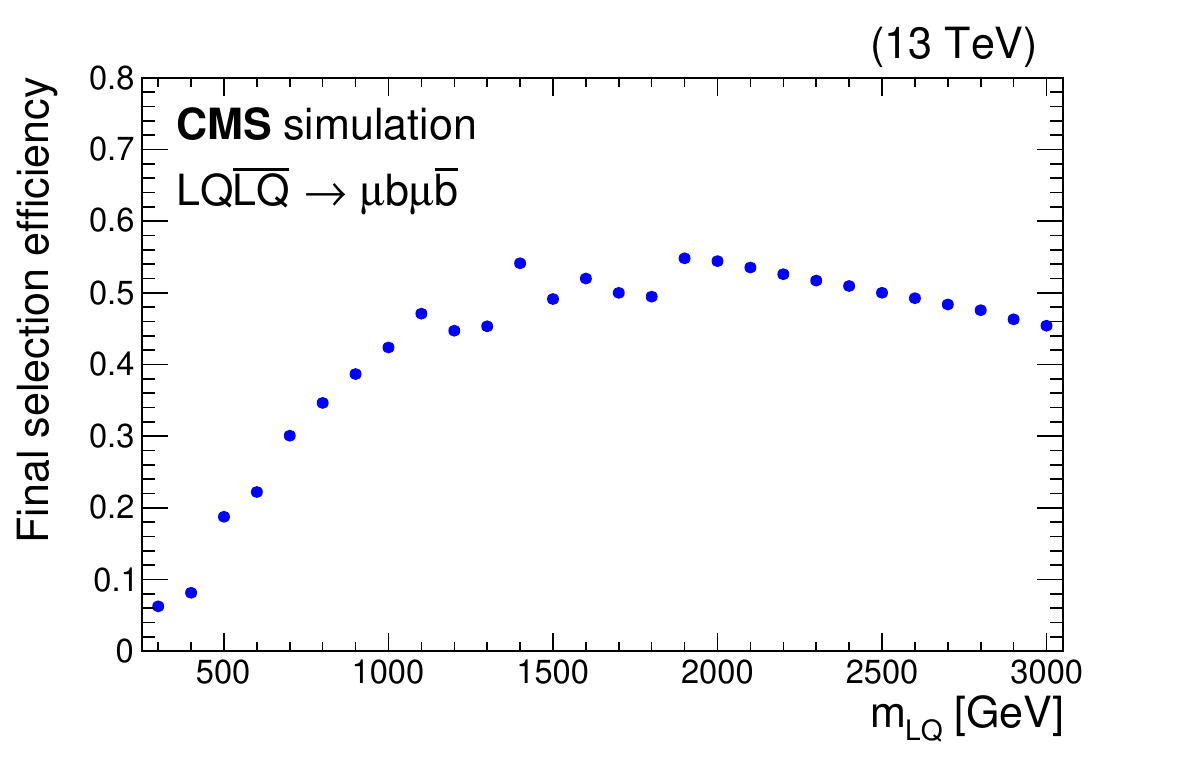}
       \caption{Total signal selection efficiency, defined as the number of events passing the final selection divided by the number of generated events. The discrete nature of the individual BDT training and final selection for each LQ candidate mass produces the observed variation in the efficiency. Relative uncertainties are less than one percent in all cases.}
          \label{fig:efficiency}
\end{figure*}

The signal-to-background separation is optimized individually for each $\mlq$ using the relevant BDT discriminant as the sensitive variable. The figure of merit in the optimization is the Punzi significance~\cite{Punzi} for a discovery potential of five standard deviations. This method is optimal for both making a discovery and for setting limits, and is valid in cases with low background event counts. A constant BDT discriminant selection is used for signal mass points $\mlq>1800\GeV$ to avoid instabilities in the optimization due to a limited event count in the simulated samples. The signal efficiency for each {\mlq} hypothesis, defined as the number of events passing final selection divided by the total number of generated events, is shown in Fig.~\ref{fig:efficiency}. The discrete nature of the final selection for each LQ candidate mass produces the observed variation in the efficiency. The background rejection ranges from 55 to 96\% and is above 90\% for all $\mlq>600\GeV$.
\begin{figure*}[htb!]
       \centering
       \includegraphics[width=0.75\textwidth]{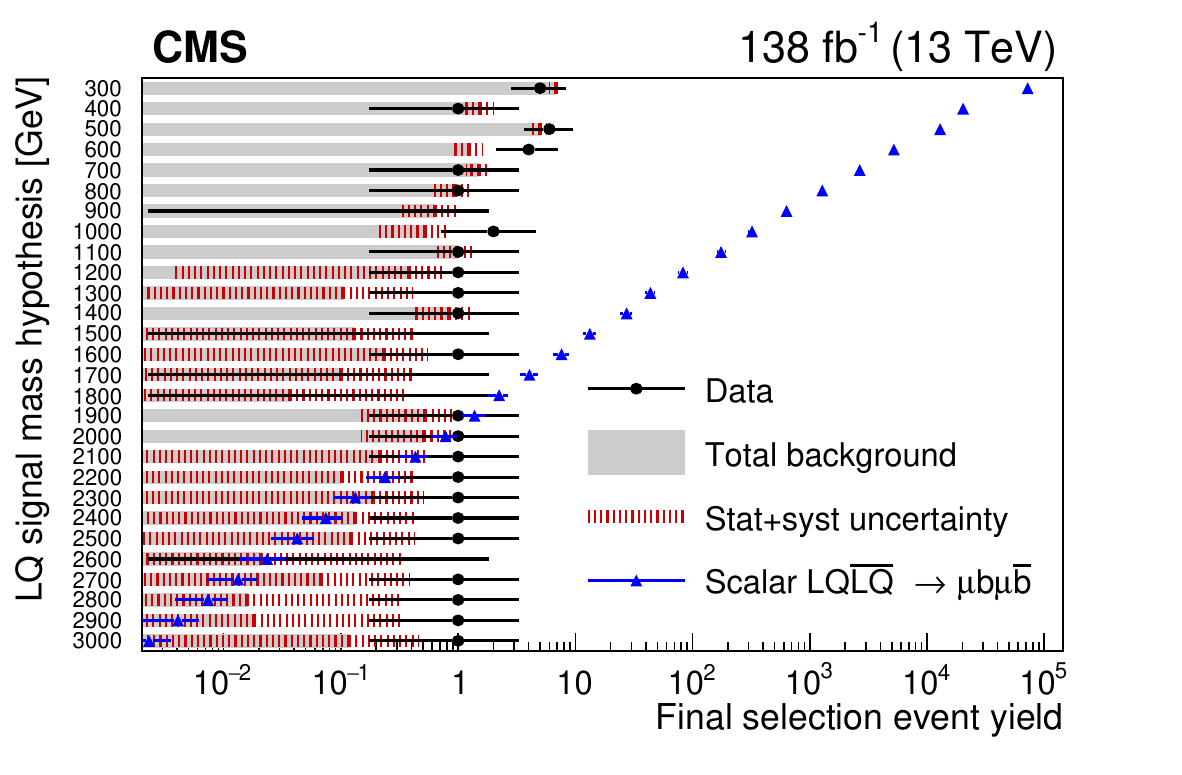}
       \caption{Data, background, and signal event yields after final selections, for each scalar {\mlq} hypothesis. Each bin on the $y$ axis represents an independent {\mlq} hypothesis. The hatched band represents the combined statistical and systematic uncertainty in the full background estimate.}
          \label{fig:finalSelTableuujj}
\end{figure*}

A detailed table of the event counts in data, expected background, and expected signal at final selection is available in the HEPData record for this analysis~\cite{hepdata}. These data are also represented visually in Fig.~\ref{fig:finalSelTableuujj}. The $x$ axis shows the final selection event yields for each of the individual {\mlq} hypotheses shown on the $y$ axis. Each bin on the $y$ axis represents an independent {\mlq} hypothesis. For example, the uppermost bin is a comparison of the yields after final selection in the $\mlq=300\GeV$ signal mass hypothesis. The hatched band represents the combined statistical and systematic uncertainty in the full background estimate.
\begin{table*}[ht!]
         \centering
                 \topcaption{Systematic uncertainties in signal efficiency and background yields in the combined 2016--2018 data set, shown as a range over all final selections (second and third columns) as well as for the $\mlq=1800\GeV$ point (rightmost two columns). The last two rows show the total systematic and statistical uncertainties in the simulated samples.}
                 \begin{scotch}{lcccc}
                                                        & \multicolumn{2}{c}{$300\leq\mlq\leq3000\GeV$} & \multicolumn{2}{c}{$\mlq=1800\GeV$} \\ 
                         Uncertainty                    & Signal (\%)   & Background (\%)  & Signal (\%)   & Background (\%)  \\ 
                         \hline
                         {\PQb} jet tagging efficiency  & 2--16          & 1--23       & 9 & 6 \\                              
                         Jet energy resolution          & $<$1           & 1--17       & $<$1 & 6 \\                           
                         Jet energy scale               & 0--2           & 1--22       & $<$1 & 9 \\                              
                         Integrated luminosity          &    4           & 0--1        & 4 & 1 \\                             
                         Muon energy resolution         & $<$1           & 0--14       & $<$0.1 & 5 \\                           
                         Muon energy scale              & 0--3           & 3--32       & $<$1 & 12 \\                             
                         Muon identification            & 2              & $<$1        & 2 & 1 \\                             
                         Muon isolation                 & $<$1           & $<$1        & $<$1 & $<$1 \\                             
                         Muon reconstruction            & 1--63          & 0--8        & 17 & 2 \\                             
                         Muon trigger                   & 0--1           & $<$1        & 1 & $<$1 \\                             
                         L1 prefire weighting           & 1              & 0--3        & 1 & 2 \\                             
                         Pileup                         & $<$1           & 0--18       & $<$1 & 4 \\                              
                         PDF                            & \NA             & 5--29       & \NA & 16 \\                              
                         Top quark {\pt} reweighting    & \NA             & 3--4        & \NA & 4 \\                              
                         \ttbar normalization           & \NA             & 1--2        & \NA & 1 \\                              
                         \ttbar shape                   & \NA             & 3--23       & \NA & 19 \\                              
                         Diboson normalization          & \NA             & 0--13       & \NA & 2 \\                              
                         Diboson shape                  & \NA             & 0--10       & \NA & 1 \\                              
                         {\zjets} normalization         & \NA             & 1--25       & \NA & 4 \\                               
                         {\zjets} shape                 & \NA             & 0--12       & \NA & 3 \\                          
                         Total syst. uncertainty        & 5--65          & 7--72       & 20 & 32 \\                      
                         Total stat. uncertainty        & 0--3           & 22--100     & $<$1 & 100 \\                  
                 \end{scotch}
                 \label{tab:syst_uujj}
\end{table*}

\subsection{Systematic uncertainties}
\label{sec:syst}
Theoretical uncertainties in the LQ signal production cross sections vary from 13 to 36\% across the $\mlq$ range of 300--2500\GeV.  They are estimated by varying the PDF eigenvectors within their uncertainties and the renormalization and factorization scales by factors of one-half and two~\cite{lqToolbox}.

Systematic uncertainties in the background yields and in the signal efficiency are calculated for each final selection by running the full analysis with separately varied detector conditions, particle momenta, or scale factors. These yields are compared to those for the nominal analysis, and the differences from nominal are used in the analysis of uncertainties.

Systematic uncertainties in the jet energy resolution~\cite{CMS:2016lmd} and muon momentum resolution~\cite{CMS:2018rym} are measured by smearing the jet and muon momenta, including high-{\pt} specific corrections for muons~\cite{Zprime}. Uncertainties due to the jet energy scale and the muon momentum scale are estimated by propagating up and down the uncertainties in the applied jet and muon momentum corrections. 

Shape uncertainties of the main background predictions are estimated by independently varying the factorization and normalization scales in the simulation by factors of one-half and two. The PDF uncertainty is estimated by varying the \textsc{NNPDF} eigenvectors within their uncertainties, following the PDF4LHC prescription~\cite{pdf-syst-1,pdf-syst-2}.

The uncertainties in the \zjets, {\ttbar}+jets, diboson, and {\ttV} background normalizations are estimated by varying the normalization scale factors described in Section~\ref{bg} up and down by their statistical uncertainties.

Other sources of systematic uncertainty include: the luminosity measurement~\cite{LUM-17-003,LUM-17-004,LUM-18-002}; muon reconstruction, identification, and isolation~\cite{CMS:2018rym}; {\PQb} tagging efficiency~\cite{BTV-16-002,CMS-DP-2023-005}; pileup~\cite{CMS:2018mlc}; trigger efficiency and prefiring (L1 trigger incorrectly assigned to an earlier bunch crossing); top {\pt} reweighting (to account for differences in top quark {\pt} spectra observed in data and simulation)~\cite{CMS:2016oae}; and track reconstruction efficiency. For LQ mass hypotheses above 1 \TeV, the systematic uncertainties in the muon reconstruction efficiency become dominant and have a significant effect on the signal samples. These large uncertainties are due to limited numbers of very high momentum muons in the data control samples used to derive the data-to-simulation correction factors.

The effects of these systematic uncertainties in signal efficiency and total background yield are shown in Table~\ref{tab:syst_uujj}. The maximum values given in Table~\ref{tab:syst_uujj} are only relevant for large values of \mlq, where the total uncertainty is dominated by the statistical uncertainty in the simulated background samples. For most values of {\mlq}, the systematic uncertainties are at the lower end of the range, as can be seen in the two rightmost columns of the table, which shows the uncertainties for the $\mlq=1800\GeV$ signal point.

\begin{figure}[htbp]
  \centering
  \includegraphics[width=\cmsFigWidth]{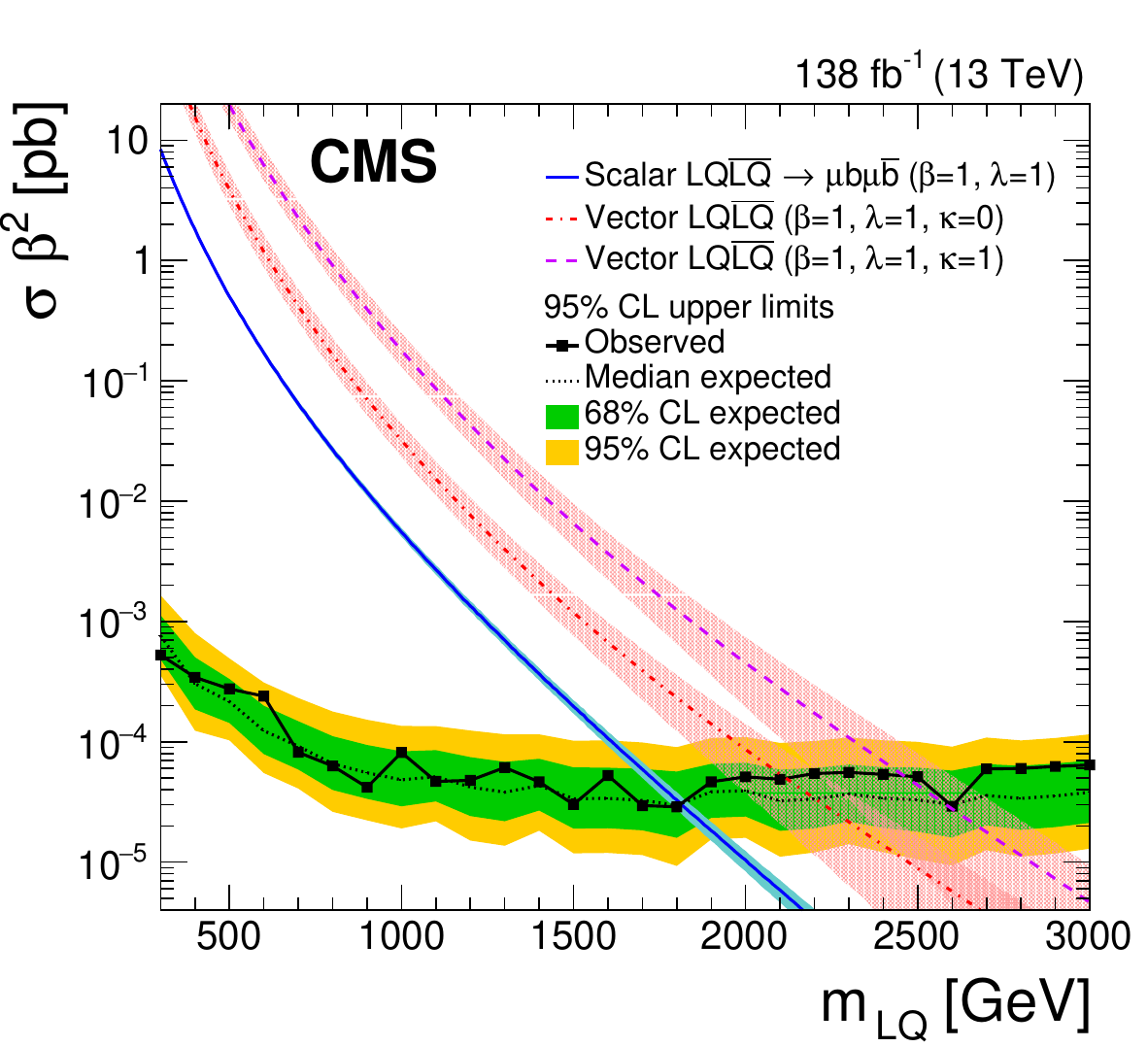}
   \caption{The expected and observed upper limits at 95\% {\CL} on the product of the LQ pair production cross section and the branching fractions $\beta^2$ as a function of {\mlq}. The black solid line represents the observed limits, the dotted line is for the median expected limits, and the inner dark-green and outer light-yellow bands are for the 68 and 95\% {\CL} intervals. The solid blue line and corresponding blue band represents the theoretical scalar LQ pair production cross sections and the uncertainties on the cross sections due to the PDF prediction and renormalization and factorization scales, respectively. Similarly, the dash-dotted (dashed) line and corresponding band represents the cross sections of theoretical vector LQ pair production and uncertainties in the minimal coupling (Yang--Mills) scenario.}
  \label{fig:limit_plot_mu}
\end{figure}
\subsection{Limit setting}

The data are compared to background predictions after the final selections have been applied. No significant excess above the predicted background is identified for any {\mlq}. Limits are set on the LQ pair production cross section as a function of {\mlq}, obtained using the asymptotic approximation~\cite{Cowan:2010js} of the modified frequentist {\CLs} approach~\cite{cls1,cls2}, which uses the ratio of the tail probabilities in the signal+background to background hypotheses. The systematic uncertainties described above are introduced as nuisance parameters in the limit setting procedure using log-normal probability functions. Uncertainties of a statistical nature are described by $\Gamma$ distributions with widths determined by the number of events in simulated samples or observed in data CRs. Individual data sets from different years are treated separately, then statistically combined into a single limit on the full data set. Most systematic uncertainties are treated as fully uncorrelated across years in the combination, with the exception of PDF, pileup, and shape uncertainties, which are fully correlated, and the integrated luminosity uncertainty, which is partially correlated.

The 95\% {\CL} upper limits on $\sigma\beta^2$ as a function of {\mlq} are shown in Fig.~\ref{fig:limit_plot_mu}, together with the NLO (LO) predictions for the scalar (vector) LQ pair production cross sections. Theoretical uncertainties in the LQ signal production cross sections are shown as a band around the signal production cross section.  Uncertainties in the vector model cross sections are larger than in the scalar model because of the LO assumptions in the interference of the helicity states of the vector model ~\cite{lqToolbox,Hammett:2015sea}. The discrete nature of the final selection for each LQ candidate mass produces the observed variation in the observed limit. By comparing the observed upper limit with the theoretical cross section values, scalar LQs with $\mlq<1810\GeV$ are excluded under the assumption of $\beta = 1$, in agreement with the median expected limit of 1810\GeV. Vector LQs with $\mlq<2120$ (2460)\GeV are excluded in the minimal coupling (Yang--Mills) scenario, under the assumption of $\beta = 1$, compared to the median expected limit of 2200 (2580)\GeV.  

\begin{figure}[htb!]
  \centering
  \includegraphics[width=\cmsFigWidth]{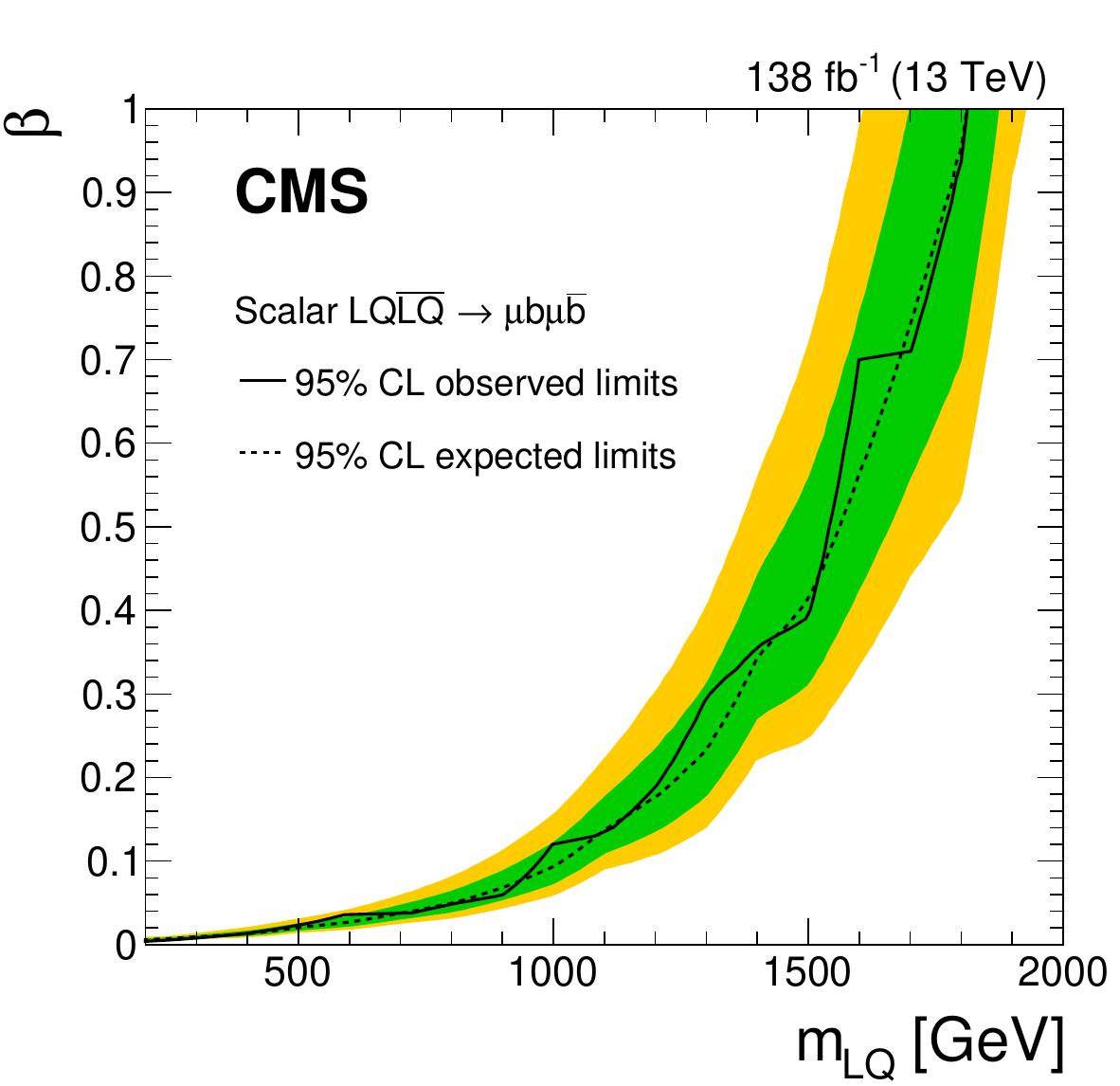}
   \caption{The expected and observed exclusion limits at 95{\%} {\CL} as a function of the leptoquark mass and the branching fraction $\beta$. The solid line represents the observed limits, the dashed line represents the median expected limits, and the inner dark-green and outer light-yellow bands represent the 68 and 95\% {\CL} intervals. The area left of the observed limit is excluded.}
  \label{fig:limit_plot_beta}
\end{figure}

Figure~\ref{fig:limit_plot_beta} shows the expected and observed exclusion limits at 95{\%} {\CL} as a function of the scalar LQ mass and $\beta$. Scalar LQs with $\mlq<1540\GeV$ are excluded for $\beta = 0.5$, compared to the median expected limit of 1560\GeV.

\section{Summary}
\label{sec:summary}
A search has been performed for pair production of leptoquarks (LQs) decaying to muons and bottom quarks using proton-proton collision data collected at $\sqrt{s} = 13\TeV$ in 2016--2018 with the CMS detector at the LHC, corresponding to an integrated luminosity of 138\fbinv. Limits are set at 95\% confidence level on the product of the scalar LQ pair production cross section and $\beta^2$, as a function of the LQ mass \mlq, where $\beta$ is the branching fraction of the LQ decaying to a muon and a bottom quark.  Scalar LQs with $\mlq<1810\GeV$ are excluded for $\beta=1$. The results are also presented as a function of $\beta$, and scalar LQs with $\mlq<1540\GeV$ are excluded for $\beta=0.5$. A further interpretation is performed with a vector LQ model, and vector LQs with $\mlq<2120$ (2460)\GeV are excluded in the minimal coupling (Yang--Mills) scenario for $\beta=1$. These represent the most stringent limits to date on these models.

\begin{acknowledgments}
We congratulate our colleagues in the CERN accelerator departments for the excellent performance of the LHC and thank the technical and administrative staffs at CERN and at other CMS institutes for their contributions to the success of the CMS effort. In addition, we gratefully acknowledge the computing centers and personnel of the Worldwide LHC Computing Grid and other centers for delivering so effectively the computing infrastructure essential to our analyses. Finally, we acknowledge the enduring support for the construction and operation of the LHC, the CMS detector, and the supporting computing infrastructure provided by the following funding agencies: SC (Armenia), BMBWF and FWF (Austria); FNRS and FWO (Belgium); CNPq, CAPES, FAPERJ, FAPERGS, and FAPESP (Brazil); MES and BNSF (Bulgaria); CERN; CAS, MoST, and NSFC (China); MINCIENCIAS (Colombia); MSES and CSF (Croatia); RIF (Cyprus); SENESCYT (Ecuador); ERC PRG, RVTT3 and TK202 (Estonia); Academy of Finland, MEC, and HIP (Finland); CEA and CNRS/IN2P3 (France); SRNSF (Georgia); BMBF, DFG, and HGF (Germany); GSRI (Greece); NKFIH (Hungary); DAE and DST (India); IPM (Iran); SFI (Ireland); INFN (Italy); MSIP and NRF (Republic of Korea); MES (Latvia); LAS (Lithuania); MOE and UM (Malaysia); BUAP, CINVESTAV, CONACYT, LNS, SEP, and UASLP-FAI (Mexico); MOS (Montenegro); MBIE (New Zealand); PAEC (Pakistan); MES and NSC (Poland); FCT (Portugal); MESTD (Serbia); MCIN/AEI and PCTI (Spain); MOSTR (Sri Lanka); Swiss Funding Agencies (Switzerland); MST (Taipei); MHESI and NSTDA (Thailand); TUBITAK and TENMAK (Turkey); NASU (Ukraine); STFC (United Kingdom); DOE and NSF (USA).
   
\hyphenation{Rachada-pisek} Individuals have received support from the Marie-Curie program and the European Research Council and Horizon 2020 Grant, contract Nos.\ 675440, 724704, 752730, 758316, 765710, 824093, and COST Action CA16108 (European Union); the Leventis Foundation; the Alfred P.\ Sloan Foundation; the Alexander von Humboldt Foundation; the Science Committee, project no. 22rl-037 (Armenia); the Belgian Federal Science Policy Office; the Fonds pour la Formation \`a la Recherche dans l'Industrie et dans l'Agriculture (FRIA-Belgium); the Agentschap voor Innovatie door Wetenschap en Technologie (IWT-Belgium); the F.R.S.-FNRS and FWO (Belgium) under the ``Excellence of Science -- EOS" -- be.h project n.\ 30820817; the Beijing Municipal Science \& Technology Commission, No. Z191100007219010 and Fundamental Research Funds for the Central Universities (China); the Ministry of Education, Youth and Sports (MEYS) of the Czech Republic; the Shota Rustaveli National Science Foundation, grant FR-22-985 (Georgia); the Deutsche Forschungsgemeinschaft (DFG), under Germany's Excellence Strategy -- EXC 2121 ``Quantum Universe" -- 390833306, and under project number 400140256 - GRK2497; the Hellenic Foundation for Research and Innovation (HFRI), Project Number 2288 (Greece); the Hungarian Academy of Sciences, the New National Excellence Program - \'UNKP, the NKFIH research grants K 124845, K 124850, K 128713, K 128786, K 129058, K 131991, K 133046, K 138136, K 143460, K 143477, 2020-2.2.1-ED-2021-00181, and TKP2021-NKTA-64 (Hungary); the Council of Science and Industrial Research, India; ICSC -- National Research Center for High Performance Computing, Big Data and Quantum Computing, funded by the EU NexGeneration program (Italy); the Latvian Council of Science; the Ministry of Education and Science, project no. 2022/WK/14, and the National Science Center, contracts Opus 2021/41/B/ST2/01369 and 2021/43/B/ST2/01552 (Poland); the Funda\c{c}\~ao para a Ci\^encia e a Tecnologia, grant CEECIND/01334/2018 (Portugal); the National Priorities Research Program by Qatar National Research Fund; MCIN/AEI/10.13039/501100011033, ERDF ``a way of making Europe", and the Programa Estatal de Fomento de la Investigaci{\'o}n Cient{\'i}fica y T{\'e}cnica de Excelencia Mar\'{\i}a de Maeztu, grant MDM-2017-0765 and Programa Severo Ochoa del Principado de Asturias (Spain); the Chulalongkorn Academic into Its 2nd Century Project Advancement Project, and the National Science, Research and Innovation Fund via the Program Management Unit for Human Resources \& Institutional Development, Research and Innovation, grant B37G660013 (Thailand); the Kavli Foundation; the Nvidia Corporation; the SuperMicro Corporation; the Welch Foundation, contract C-1845; and the Weston Havens Foundation (USA).
\end{acknowledgments}

\bibliography{auto_generated}
\cleardoublepage \appendix\section{The CMS Collaboration \label{app:collab}}\begin{sloppypar}\hyphenpenalty=5000\widowpenalty=500\clubpenalty=5000
\cmsinstitute{Yerevan Physics Institute, Yerevan, Armenia}
{\tolerance=6000
A.~Hayrapetyan, A.~Tumasyan\cmsAuthorMark{1}\cmsorcid{0009-0000-0684-6742}
\par}
\cmsinstitute{Institut f\"{u}r Hochenergiephysik, Vienna, Austria}
{\tolerance=6000
W.~Adam\cmsorcid{0000-0001-9099-4341}, J.W.~Andrejkovic, T.~Bergauer\cmsorcid{0000-0002-5786-0293}, S.~Chatterjee\cmsorcid{0000-0003-2660-0349}, K.~Damanakis\cmsorcid{0000-0001-5389-2872}, M.~Dragicevic\cmsorcid{0000-0003-1967-6783}, P.S.~Hussain\cmsorcid{0000-0002-4825-5278}, M.~Jeitler\cmsAuthorMark{2}\cmsorcid{0000-0002-5141-9560}, N.~Krammer\cmsorcid{0000-0002-0548-0985}, A.~Li\cmsorcid{0000-0002-4547-116X}, D.~Liko\cmsorcid{0000-0002-3380-473X}, I.~Mikulec\cmsorcid{0000-0003-0385-2746}, J.~Schieck\cmsAuthorMark{2}\cmsorcid{0000-0002-1058-8093}, R.~Sch\"{o}fbeck\cmsorcid{0000-0002-2332-8784}, D.~Schwarz\cmsorcid{0000-0002-3821-7331}, M.~Sonawane\cmsorcid{0000-0003-0510-7010}, S.~Templ\cmsorcid{0000-0003-3137-5692}, W.~Waltenberger\cmsorcid{0000-0002-6215-7228}, C.-E.~Wulz\cmsAuthorMark{2}\cmsorcid{0000-0001-9226-5812}
% ...
\par}
\vskip\cmsinstskip
\dag:~Deceased\\
$^{1}$Also at Yerevan State University, Yerevan, Armenia\\
$^{2}$Also at TU Wien, Vienna, Austria\\
% ...
\end{sloppypar}
\end{document}